\documentclass[twocolumn]{aa}
\usepackage{graphicx}
\usepackage{txfonts}
\begin{document}
   \title{Submillimeter Emission from Water in the W3 Region\thanks{Based 
on observations
with Odin, a Swedish-led satellite project funded jointly by the
Swedish National Space Board (SNSB), the Canadian Space Agency
(CSA), the National Technology Agency of Finland (Tekes),
and Centre National d'Etudes Spatiales (CNES). The Swedish Space
Corporation was the industrial prime contractor and is also responsible for
the satellite operation.}
}
   \author{C. D. Wilson,
          \inst{1}
          A. Mason,\inst{1}
          E. Gregersen,\inst{1}
A. O. H. Olofsson,\inst{2}
P. Bergman,\inst{2}
R. Booth,\inst{2}
N. Boudet,\inst{3} 
V. Buat,\inst{4} 
C. L. Curry,\inst{5} 
P. Encrenaz,\inst{6} 
E. Falgarone,\inst{7} 
P. Feldman,\inst{8} 
M. Fich,\inst{5} 
H. G. Floren,\inst{9}  
U. Frisk,\inst{10} 
M. Gerin,\inst{7} 
J. Harju,\inst{11} 
T. Hasegawa,\inst{12} 
A. Hjalmarson,\inst{2}
M. Juvela,\inst{11}
S. Kwok,\inst{12} 
B. Larsson,\inst{9} 
A. Lecacheux,\inst{13} 
T. Liljestrom,\inst{14} 
R. Liseau,\inst{9}
K. Mattila,\inst{11}
G. Mitchell,\inst{15} 
L. Nordh,\inst{16} 
M. Olberg,\inst{2}
G. Olofsson,\inst{9}
L. Pagani,\inst{6}
R. Plume,\inst{12}
I. Ristorcelli,\inst{3} 
Aa. Sandqvist,\inst{9}
G. Serra\inst{3}
N. Tothill, \inst{15}
K. Volk,\inst{12},
          \and
F. von Scheele\inst{10}
          }

\authorrunning{C. D. Wilson et al.}

   \offprints{C. D. Wilson}

   \institute{Department of Physics \& Astronomy, McMaster University,
	Hamilton, Ontario L8S 4M1 Canada\\
              \email{wilson@physics.mcmaster.ca} 
         \and
	Onsala Space Observatory, SE-439 92, Onsala Sweden 
\and
CESR, 9 Avenue du Colonel Roche, B.P. 4346, 31029 Toulouse, France 
\and
Laboratoire d'Astronomie Spatiale, BP 8, 13376 Marseille, CEDEX 12, 
France 
\and
Department of Physics, University of Waterloo, Waterloo ON N2L 3G1, 
Canada 
\and
LERMA \& FRE du CNRS, Observatoire de Paris, 61, A. de
l'Observatoire, 75140 Paris, France 
\and
LERMA \& FRE 2460 du CNRS, Ecole Normale Superieure, 24 rue Lhomond, 75005 Paris, France 
\and
Herzberg Institute of Astrophysics, National Research Council, 5071 
West Saanich Road, Victoria B.C. V8X 4M6 Canada 
\and
Stockholm Observatory, SCFAB, Roslagstullsbacken 21, SE-106 91 Stockholm, 
Sweden 
\and
Swedish Space Corporation, P.O. Box 4207, SE-171 04 Solna, Sweden 
\and
Observatory, P. O. Box 14, University of Helsinki, 00014 Helsinki, 
Finland 
\and
Department of Physics and Astronomy, University of Calgary, Calgary AB T2N 1N4,
Canada 
\and
LESIA, Observatoire de Paris, Section de Meudon, 5, Place Jules Janssen,
92195 MEUDON CEDEX, France 
\and
Metsahovi Radio Observatory, Helsinki University of Technology, Otakaari 5A, 
FIN-02150 Espoo, Finland 
\and
Department of Astronomy and Physics, Saint Mary's University, Halifax, Nova 
Scotia B3H 3C3, Canada 
\and
Swedish National Space Board, Box 4006, SE-171 04 Solna, Sweden 
             }

   \date{January 21, 2003}

   \abstract{Using the Odin satellite,
we have mapped the submillimeter emission from the $1_{10}-1_{01}$ transition
of {\it ortho}-water in the W3 star-forming region. A
$5\arcmin\times 5\arcmin$ map of the W3 IRS4 and W3 IRS5 region reveals
strong water lines at half the positions in the map.
The relative strength of the Odin lines compared to previous observations
by SWAS suggests that we are seeing water emission from an
extended region. 
Across much of the map the lines are double-peaked, with an absorption
feature at --39 km s$^{-1}$; however, some positions in the map
show a single strong line at --43 km s$^{-1}$.
We interpret the double-peaked lines as arising from optically
thick, self-absorbed water emission near the W3 IRS5, while the
narrower blue-shifted lines originate in 
emission near W3 IRS4. In this model, the unusual appearance of the 
spectral lines across the map results from a coincidental agreement
in velocity between  the emission  near W3 IRS4 and the
blue peak of the more complex lines near W3 IRS5.
The strength of the water 
lines near W3 IRS4 suggests we may be seeing water emission enhanced in
a 
photon-dominated region.
   }

   \maketitle
%

\section{Introduction}

   \begin{figure*}
   \centering
   \includegraphics[angle=-90,width=16.1cm]{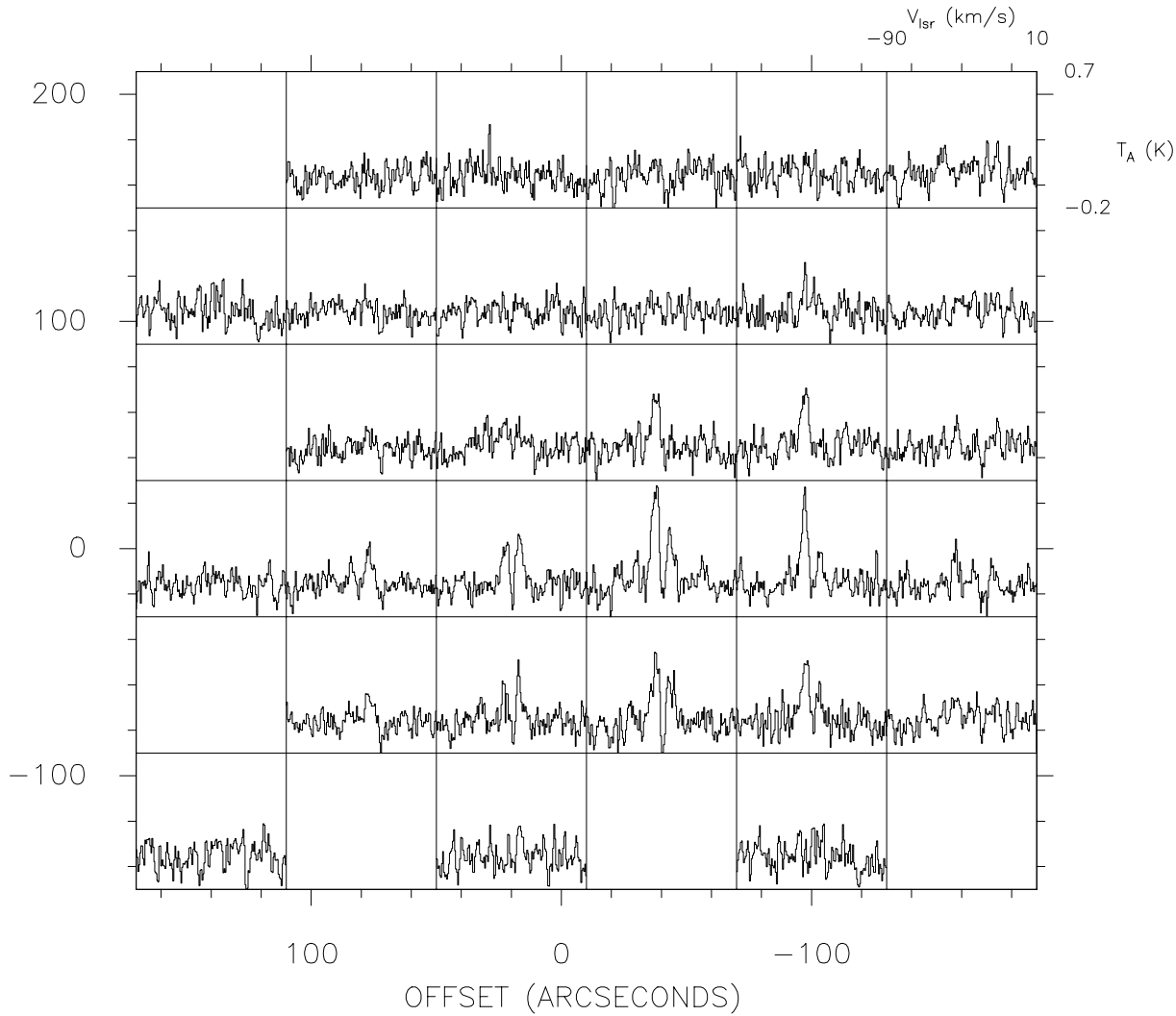}
   \caption{Individual water spectra observed towards W3 Main.
The (0,0) position is located 
at 02:25:40.5 +62:06:20 (J2000), slightly north of W3 IRS5.
The map was obtained on a 1$^\prime$ grid except for the
extreme eastern and southern edges, which were observed on
a 2$^\prime$ grid. (For ease of display, spectra on these edges
are plotted roughly (20$^{\prime\prime}$,20$^{\prime\prime}$)
from their true position; the true locations of
these spectra are shown in Fig.~\ref{Fig_int}.)
We estimate the
absolute pointing accuracy of the map to be $\sim 15^{\prime\prime}$.
}
              \label{Fig_grid}
    \end{figure*}

Water has long been thought to be a potentially important coolant
in the molecular interstellar medium (Goldsmith \& Langer
\cite{Goldsmith}). 
However, recent SWAS observations have shown its gas phase abundance 
to be significantly lower than expected, a fact which is
most likely attributable to freezing out of molecules onto
grains (i.e. Bergin et al. \cite{Bergin}).
The abundance of water shows significant variations from one region
to another, with abundances in hot cores and shock-heated outflows 
at least 100 times larger than the abundance in the colder
material that dominates in molecular clouds
(Snell et al. \cite{Snella}, Neufeld et al.
\cite{Neufeld}). In this paper, we present
a small map made with the Odin satellite (cf. Frisk et al.
\cite{Frisk}, Hjalmarson et al.
\cite{Hjalmarson}, Nordh et al. \cite{Nordh}) 
of extended water emission from the ambient molecular
gas in W3 which reveals intriguing variations in the line
shape across the 
region.

The W3 molecular cloud lies at a distance of
roughly 2.3 kpc and contains two well-studied regions of
massive star formation, W3 Main and W3(OH). W3 Main
contains an infrared cluster (Megeath et al. \cite{Megeath}), which
includes the two luminous
infrared sources W3 IRS5 and IRS4 (Wynn-Williams, Becklin, \&
Neugebauer \cite{Wynn}), a number of
extremely compact HII regions (Tieftrunk et al.
\cite{Tieftrunk1997}), and a powerful outflow
(Bally \& Lada \cite{Bally}; Hasegawa et al. \cite{Hase}). 
W3 IRS5 contains numerous water masers which
are predominantly associated with two outflows (Imai
et al. \cite{Imai}). 
Helmich et al. (\cite{Helmich}) find that W3 IRS5 is in
an earlier evolutionary stage than W3 IRS4,
which may be a photon dominated region on the far side of
the molecular core (Helmich \& van Dishoeck \cite{h97}).
The molecular gas in W3 Main is organized into three cores;
one core coincides with W3 IRS5, a second core lies just
south of W3 IRS4 (i.e. Tieftrunk et al. \cite{Tieftrunk1995}), 
and a third core lies south-east
of W3 IRS5 (Tieftrunk et al. \cite{Tieftrunk}). 

Snell et al. (\cite{Snella}) detected the 557 GHz
line of {\it ortho}-water towards W3 and other molecular cloud cores
with the SWAS satellite and estimate water abundances 
between $6\times 10^{-10}$ and $1\times 10^{-8}$. These abundances
are at least 100 times smaller than the water abundance
estimated using ISO observations of higher energy lines from
the very compact hot cores found in some of these
regions (i.e. van Dishoeck \& Helmick
\cite{vD1996}). In W3, Snell et al. (\cite{Snella}) find the strong
absorption feature in the H$_2$O spectrum coincides
with the peak of the $^{13}$CO J=5--4 emission line,
which indicates that water emission in W3 is self-absorbed.
In this paper, we present a $5^\prime \times 5^\prime$ fully-sampled map of 
the water emission from the W3 region with a factor of two better
resolution than the SWAS detection. Our map reveals intriguing
variations in the shape and strength of the water line 
(Fig.~\ref{Fig_grid}) and suggests that multiple sources and/or extended
components contribute to the water emission from this 
star-forming region.


\section{Observations}

We made a deep integration towards W3 IRS5 with Odin from 2002 Jan 25 
to 2002 Jan 28 with a total on-source integration time of 380 minutes. During
this period, we also obtained spectra of a 9 point
map with 2$^\prime$ spacing and typical integration times of
31 to 37 minutes per point. We subsequently made a 25 point map with
1$^\prime$ spacing from 2002 Mar 11 to 2002 Mar 19. 
The typical integration time was 36 minutes per
point, with a minimum
time of 33 minutes and a maximum time of 58 minutes.
All coordinates in this paper have been corrected for
a global pointing offset of (35$^{\prime\prime}$,$-10^{\prime\prime}$)
in January 2002 and (34$^{\prime\prime}$,13$^{\prime\prime}$)
in March 2002, as
determined from a map of Jupiter made in April 2002.
The Odin beam at this frequency is $2.1^\prime$ and
the typical system temperature was 3300 K (SSB). 
For all observations of W3, the spectrometer
was configured with the acousto-optic spectrometer (AOS) and one
autocorrelator observing the 556.936 GHz line of {\it ortho}-H$_2$O using two
independent receivers. 
All observations were obtained in sky-switching mode and observations
of a reference off position 30$^\prime$ east of W3 were obtained as well.

\begin{figure}
   \centering
   \includegraphics[angle=-90,width=8.5cm]{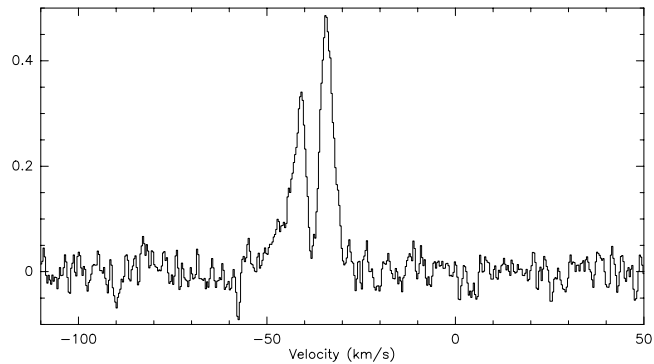}
      \caption{Deep H$_2$O integration towards W3 IRS5.
The spectrum was obtained  at 
02:25:40.6 +62:05:57 (J2000).
The blue peak is $0.48\pm 0.02$ K (T$_A^*$) and the red peak
is 0.34 K at a resolution of 
1 MHz (0.54 km s$^{-1}$).
              }
         \label{Fig_deep}
   \end{figure}

The AOS data were reduced using CLASS. To identify
individual spectra which were contaminated by the Earth's atmosphere,
we measured 
the average emission in a 60 km s$^{-1}$
region around 0 km s$^{-1}$  and removed spectra
that were significantly above or below the average value.
Spectra that were mis-tuned or that had
anomalous system temperatures were also removed.
The spectra for the off position
were averaged and then smoothed with a gaussian function
with a velocity full-width half-maximum of 1 km s$^{-1}$. 
This smoothed  spectrum  was
subtracted from each individual source spectrum 
to correct for the intrinsic
fast ripple seen in the AOS spectra in
sky-switching mode (Hjalmarson et al. \cite{Hjalmarson}). 
Finally, all spectra  within $20^{\prime\prime}$ of a
given position were averaged together. 
The spectra were then hanning smoothed 
and first to third order baselines were fit over the
region shown in Fig.~\ref{Fig_deep}  and 
excluding a window around the line
from --60 to --20 km s$^{-1}$ for spectra with
detections of the water line. The deep integration on W3 IRS5 
shown in Fig.~\ref{Fig_deep} has
not been hanning smoothed and has had a fourth order baseline removed.
The rms noise in the deep spectrum is 0.024 K at a resolution of 
1 MHz (0.54 km s$^{-1}$), 
while the rms noise in the individual spectra in the map ranges
from 0.056 to 
0.085 K at a 
resolution of 1.24 MHz (0.67 km s$^{-1}$).

   \begin{figure}
   \centering
   \includegraphics[width=7.4cm]{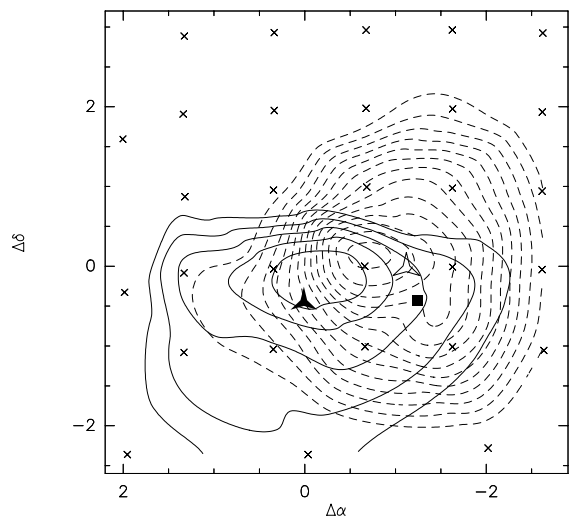}
      \caption{Integrated intensity of the red 
(--39 to -- 28 km s$^{-1}$, solid contours) 
and blue (--52 to --39 km s$^{-1}$, dashed contours) 
components of the water lines in W3. 
Contours start at 0.75 K km s$^{-1}$ and increase 
in steps of 0.25 K km s$^{-1}$ ($\sim 1\sigma$).
The positions of W3 IRS5 and IRS4 are marked
with filled and open pinwheels, respectively, and 
the SWAS pointing is marked with the filled square.
              }
         \label{Fig_int}
   \end{figure}

\section{Extended water emission towards W3 Main}

The relative strengths of the SWAS (Snell et al. \cite{Snella}) and Odin 
H$_2$O detections suggest that the water emission in this region
of W3 does not originate in a single, very compact source. The beam
area of SWAS ($3.3^\prime \times 4.5^\prime$) 
is 3.3 times larger than that of Odin at this frequency,
yet the peak line strength measured by Odin in the W3 map 
is only 2.1 times stronger than the line measured by SWAS. 
In addition, the water emission observed with Odin appears significantly 
extended relative to the beam, with the red and blue components
having different spatial distributions
(Fig.~\ref{Fig_int}). The observed full-width half-maximum diameter
is $2.6^\prime \times 2.4^\prime$, which suggests a true
diameter of $1.4^\prime\times \sim 1.2^\prime$. Individually,
the red and blue components of the line are also significantly larger
than the beam.
Either the water emission in this region is significantly extended,
or there are two or more strong point sources contained within
the SWAS and Odin beams.

Two obvious candidates for point sources of water emission are W3
IRS5 and IRS4, which are separated by $\sim 70^{\prime\prime}$.
However, the pointings of the SWAS spectrum and the strongest Odin spectrum
relative to the locations of the two sources
are such that the expected line strength ratio between the two satellites
would still be about 3 if both sources are point-like. 
If the water emission seen in this region is associated with
W3 IRS5 and IRS4, one or both of the sources must be significantly
extended relative to the Odin beam.
C$^{18}$O mapping by Tieftrunk et al.
(\cite{Tieftrunk1995}) reveals two molecular cores in this region,
one $20^{\prime\prime}\times 20^{\prime\prime}$
associated with W3 IRS5 and a second $40^{\prime\prime} 
\times 30^{\prime\prime}$ core which lies 
slightly to the south of W3 IRS4 and shares the same
central velocity. It seems likely that these molecular cores may
account at least in part for the extent of the water
emission in the W3 region.

   \begin{figure}
   \centering
   \includegraphics[angle=-90,width=8.5cm]{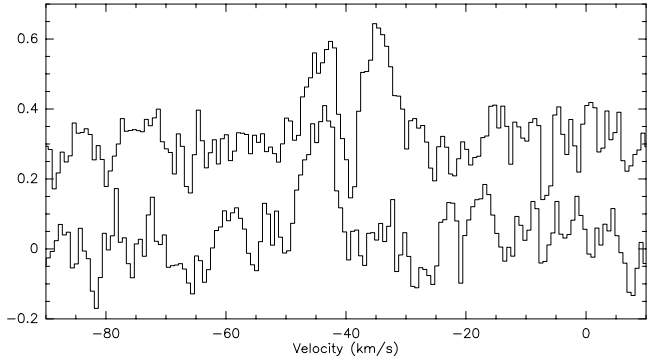}
      \caption{Comparison of water spectra near W3 IRS5 (top) and W3 IRS4
(bottom). The vertical
scale is --0.2 to 0.7 K and the W3 IRS5 spectrum has
been offset by 0.3 K for clarity.
Note the alignment in velocity between the blue wing of the W3 IRS5
spectrum and the peak of the W3 IRS4 spectrum. 
              }
         \label{Fig_comp}
   \end{figure}

The center of the absorption feature in the H$_2$O spectra lies
at $\sim -39$ km s$^{-1}$ across the map (see Fig.~\ref{Fig_deep}).
The location of this absorption feature matches the central velocity of
NH$_3$ emission from W3 IRS5 quite well, and so towards this
region we are likely seeing strongly self-absorbed water lines,
similar to the CO J=3--2 and J=6--5 lines observed
by Hasegawa et al. (\cite{Hase}).
However, the velocity towards W3 IRS4 and the molecular core
just south of it is significantly different
(--42.5 km s$^{-1}$, Tieftrunk
et al. \cite{Tieftrunk}, Hasegawa et al. \cite{Hase}). 
Thus, it is difficult 
to understand the constant velocity of the --39 km s$^{-1}$ absorption feature 
across the map given the velocity shift of 3-4 km s$^{-1}$ that is seen
in optically thin tracers such as C$^{18}$O and NH$_3$
moving from west (near W3 IRS4) to east (W3 IRS5) across the region
(Tieftrunk et al. \cite{Tieftrunk1995}, \cite{Tieftrunk}). 

The most likely possibility is that we are seeing water
emission from two distinct regions in our map, one associated with
W3 IRS5 and the second associated with W3 IRS4 and/or its molecular
core. In this model,
it is the {\it coincidental} agreement in velocity between the 
single-peaked line from W3 IRS4 
and the blue-shifted peak of the self-absorbed
spectrum from the W3 IRS5 region that produces the constant
velocity of the absorption feature at --39 km s$^{-1}$.
This interpretation is illustrated in Fig.~\ref{Fig_comp}, which compares
the observed spectra near W3 IRS5 and W3 IRS4.
The W3 IRS5 region contains at least two outflows
(Hasegawa et al. \cite{Hase}, Imai et al. \cite{Imai}),
which could be partly responsible for the prominence
of the red-shifted emission peak. The second region is associated with 
W3 IRS4 itself 
(and its associated photon dominated region) 
and/or the
extended molecular core near W3 IRS4 (Tieftrunk et al. \cite{Tieftrunk1995},
\cite{Tieftrunk}). If water emission from this region is not self-absorbed,
we would expect the emission to peak at a velocity of --42 to --43 km s$^{-1}$,
in reasonable agreement with our observations. 

We have estimated the H$_2$O abundance towards W3 IRS5 using
the formula given by Snell et al. (\cite{Snellb}). We use
a density of $1\times 10^6$ cm$^{-3}$ (Snell et al. \cite{Snella}) and
adopt a temperature of 40 K. We estimate an H$_2$ column density
of $5\times 10^{22}$ cm$^{-2}$ inside the Odin beam from the
mass of 1900 M$_\odot$ derived for the W3 core (Tieftrunk et al. 
\cite{Tieftrunk1995}, scaled to a temperature of 40 K). The
integrated intensity of $3.9\pm 0.1$ K km s$^{-1}$ 
(Fig.~\ref{Fig_deep}) results in an {\it ortho}-H$_2$O abundance 
of $2\times 10^{-9}$, in excellent agreement with the 
results from SWAS (Snell et al. \cite{Snella}).
We will present
a more detailed analysis of the W3 region, including follow-up
H$_2^{18}$O and NH$_3$ observations, 
in a future paper.

\begin{acknowledgements}
Generous financial support from the Research Councils
and Space Agencies in Canada, Finland, France, and Sweden
is gratefully acknowledged.
\end{acknowledgements}

\end{document}